\documentclass[aps,prx,twocolumn,superscriptaddress,amsmath,amssymb,floatfix]{revtex4-2}

\usepackage{graphicx}
\usepackage{dcolumn}
\usepackage{bm}
\usepackage{hyperref}

\usepackage[caption = false]{subfig}
\usepackage{csquotes}
\usepackage{amsmath}

\usepackage[normalem]{ulem}
\usepackage{xcolor}

\begin{document}


\title{Secondary Electron-Only Reconnection Driven by Large Scale Ion-Coupled Reconnection and Electron Kelvin-Helmholtz Instabilities in Hybrid Simulations of Solar Wind Turbulence}

\author{Joaquín Espinoza-Troni}
\affiliation{Departamento de Física, Facultad de Ciencias,
Universidad de Chile, Santiago, Chile}
\email{joaquinun@gmail.com}

\author{Giuseppe Arrò}
\affiliation{Department of Physics, University of Wisconsin-Madison, Madison, WI 53706, USA}

\author{Francesco Califano}
\affiliation{Dipartimento di Fisica \enquote{E. Fermi}, Università di Pisa, Pisa, Italy}

\author{Julia E. Stawarz}
\affiliation{School of Engineering, Physics, and Mathematics, Northumbria University, Newcastle upon Tyne, NE1 8ST, UK}

\author{Pablo S. Moya}
\affiliation{Departamento de Física, Facultad de Ciencias,
Universidad de Chile, Santiago, Chile}


\begin{abstract}
Electron-only reconnection (EREC) is a magnetic reconnection regime occurring within subion-scale current sheets (CSs), exhibiting only electron jets, without any ion outflows. EREC has been first observed in the Earth's magnetosheath, where its occurrence is linked to the small correlation length of magnetic fluctuations, limiting the growth of CSs to very large scales. On the other hand, the development of EREC in open systems with large magnetic correlation lengths, such as the solar wind (SW), remains an open question. To address this problem, we employ a large-scale 2D hybrid simulation with finite electron inertia, investigating the development of EREC driven by turbulence. By injecting energy at very large scales, we allow EREC to develop spontaneously due to the turbulent cascade, without any external small-scale forcing or imposed constraints on the turbulence correlation length. We find that EREC develops in our simulation via two distinct turbulence-driven mechanisms: (1) secondary EREC induced by the interaction of plasmoids in the outflows of large-scale ion-coupled reconnection; (2) EREC directly driven at subion scales by the electron Kelvin-Helmholtz instability in small-scale velocity shears. Furthermore, we perform a statistical analysis of CSs using the machine-learning clustering algorithm HDBSCAN, showing that subion-scale CSs capable of hosting EREC are dominant in our simulation. Our results suggest that EREC could occur even in large-scale space and astrophysical systems, like the SW, driven by secondary turbulent processes, potentially playing a key role in dissipating energy at kinetic scales. 
\end{abstract}



\maketitle

\section{Introduction} 

Magnetic reconnection is a process that occurs in magnetized plasmas, altering the magnetic field topology, typically converting magnetic energy into plasma kinetic and internal energies \citep[see e.g.][for a review]{biskamp_2000}. On top of being the driver of several extreme space and astrophysical phenomena \citep{zweibel2009magnetic,yamada2010magnetic}, reconnection is also a fundamental element of turbulence \citep{servidio2009magnetic}. Plasma turbulence naturally produces intermittent current sheets (CSs) where magnetic reconnection occurs \citep{zhdankin2013statistical}, supporting the energy cascade \citep{franci2017magnetic,cerri2017reconnection}, and dissipation \citep{sundkvist2007dissipation,servidio2012local,franci2022anisotropic}. Typically, turbulence-driven reconnection occurs in small electron-scale diffusion regions embedded within larger ion-scale CSs \citep[see][and references therein]{zweibel2016perspectives}. In this case, both ions and electrons are accelerated by reconnection, producing fast divergent bidirectional jets flowing out of the reconnection region, as observed in CSs within the turbulent solar wind (SW) \citep{gosling2005direct,eriksson2022characteristics}. In addition to the standard reconnection scenario, Magnetospheric Multiscale (MMS) \citep{burch2016magnetospheric} observations in the Earth's turbulent magnetosheath have revealed peculiar reconnection events where only electrons appear to be accelerated. These unusual reconnection events occur in electron-scale CSs, where only electron jets are observed, with no accompanying ion outflows \citep{phan2018electron}. This novel reconnection regime has been named \enquote{electron-only reconnection} (EREC), to distinguish it from standard \enquote{ion-coupled reconnection} (IREC). 

The observation of EREC events in the Earth's magnetosheath raises the question of how turbulence can produce electron-scale CSs, bypassing ion scales. This is a nontrivial problem since turbulence typically involves a continuous energy transfer from magnetohydrodynamic to kinetic scales, and turbulent fluctuations should couple to ions first, before reaching electron scales. Numerical studies suggest that the dynamics of reconnection in turbulence is determined by the correlation length of turbulent fluctuations, which typically corresponds to the energy-injection scale \citep{Pyakurel_etal_2019}. Specifically, when energy is injected at scales of about $10\,d_i$ or less (with $d_i$ being the ion inertial length), ions decouple from the magnetic field dynamics, and reconnection is driven only by electrons, while for larger injection scales, turbulence induces IREC \citep{califano2020electron,arro2020statistical,granier2024electron}. This scenario is consistent with the occurrence of EREC in the Earth's magnetosheath, where the correlation length of turbulent fluctuations ranges from approximately $10\,d_i$ in the dayside, to about $40\,d_i$ in the flanks \citep{stawarz2022turbulence,lapenta2025lagrangian}. In this environment, observations show a tendency for EREC to dominate over IREC in regions with smaller correlation lengths \citep{stawarz2019properties,stawarz2022turbulence}. Nonetheless, it remains unclear whether EREC can also develop in environments such as the SW, where the correlation length of turbulence can reach hundreds or even thousands of $d_i$ \citep{bandyopadhyay2020situ}. Numerical \citep{lapenta2015secondary} and observational \citep{zhou2021observations} studies have shown that large-scale magnetotail reconnection drives secondary EREC in its turbulent outflows, and similar mechanisms may also be at play in the SW, driven by local IREC events \citep{zhong2021three,lapenta2022formation}, or by shear flow instabilities \citep{karimabadi2013coherent,arro2023generation,Espinoza-Troni_etal_2025}, producing subion scale structures. Furthermore, recent studies have highlighted the role of EREC in distributing the dissipated magnetic energy between ion and electrons in turbulent environments \citep{Stawarz_etal_2024}. Hence, understanding the occurrence of EREC in turbulent systems with large correlation lengths, such as the SW, is relevant to comprehend the nature of non-linear interactions at kinetic scales, the energy partition between species, and the broader role of reconnection as a collisionless dissipation mechanism within these specific environments.

In this work, we use a large-scale high-resolution 2D Hybrid Vlasov-Maxwell simulation of plasma turbulence to investigate the spontaneous development of EREC in a SW-like environment. Unlike previous work, we initialize turbulence by injecting energy at scales much larger than ion scales, ensuring that EREC events develop naturally as a consequence of the large-scale turbulent cascade, without the need to force turbulence directly near ion scales. We identify two distinct turbulence-driven mechanisms that spontaneously produce EREC from large-scale fluctuations in our simulation. In the first mechanism, turbulence creates large-scale elongated CSs in which IREC occurs, producing ion-scale plasmoid chains; these plasmoids are advected in the outflows of IREC events, where they interact and merge, triggering small-scale EREC events. In the second mechanism, the large-scale turbulent dynamics first produce thin, elongated electron-velocity shears that are unstable to the electron Kelvin-Helmholtz instability (EKHI); subsequently, EKHI-generated subion-scale electron vortices locally bend magnetic field lines, inducing EREC. We find that the two mechanisms described above are ubiquitous in the simulation and might contribute to the production of EREC events in the turbulent SW. Furthermore, to assess the global occurrence and relevance of EREC in our simulations, we complement our study with a statistical analysis of CSs using the HDBSCAN clustering algorithm. We find that the vast majority of CSs produced by turbulence in our simulation have thicknesses smaller than $d_i$ and are associated with significant energy conversion, possibly due to EREC. These results highlight that EREC is a common feature of plasma turbulence, not only in relatively small systems such as the Earth's magnetosheath, but also in the large-scale turbulent SW.    

\section{Simulation setup}

Our simulation was performed using the Eulerian Hybrid Vlasov-Maxwell code \citep{valentini2007hybrid}, with kinetic ions and finite-mass fluid electrons. The electron dynamics are encoded in the generalized Ohm's law 
\begin{equation}
\begin{gathered}
\textbf{E} - \frac{d^2_e}{n}\nabla^2\textbf{E} = - \left( \textbf{u}_i \times \textbf{B} \right) + \frac{1}{n}\left( \textbf{J} \times \textbf{B} \right) - \frac{1}{n} \nabla P_e +
\\[5pt]
+ \frac{d^2_e}{n}\nabla\cdot\left( \textbf{u}_i \textbf{J} + \textbf{J}\textbf{u}_i - \frac{\textbf{J}\textbf{J}}{n} \right),
\end{gathered}
\end{equation}
where $\textbf{E}$ and $\textbf{B}$ are the electric and magnetic fields, $\textbf{J}\!=\!\nabla\times\textbf{B}$, $\textbf{u}_i$ and $n$ are the ion fluid velocity and density (quasi-neutrality is assumed), $P_e$ is the electron pressure, and $d_e$ is the electron inertial length. Electron inertia effects included in our hybrid model allow electrons to demagnetize at scales of order $d_e$, despite the absence of electron kinetic effects \citep{valentini2007hybrid, Goldman_etal_2016}. 

We consider a 2D square periodic domain of size $L\!=\!100\,\pi\,d_i$, sampled by a uniform grid with $3072^2$ points. Velocity space is sampled by a cubic uniform mesh with $51^3$ points, ranging from $-5\,v_{th,i}$ to $5\,v_{th,i}$ in each direction (with $v_{th,i}$ being the initial ion thermal speed). Ions are initially Maxwellian, with uniform density $n_0$, zero mean velocity, and isotropic temperature $T_i$. The ion beta is $\beta_i\!=\!1$ and electrons are isothermal with temperature $T_e$. The ion-to-electron mass and temperature ratios are $m_i/m_e\!=\!100$ and $T_i/T_e\!=\!1$. Turbulence is freely-decaying, initialized by adding random-phase, isotropic magnetic perturbations to a uniform out-of-plane guide field $\textbf{B}_0\!=\!B_0\hat{\textbf{z}}$. Magnetic fluctuations have wavenumbers $k$ uniformly distributed in the range $1\!\leqslant\!k/k_0\!\leqslant\!6$ (with $k_0\!=\!2\pi/L$), and root mean square (rms) amplitude $\delta B_{rms}/B_0\!\simeq\!0.28$. This initialization implies an injection scale of about $50\,d_i$, much larger than ion scales. This choice for the injection scale allows us to consider a turbulent system that is large enough as compared to the Earth's magnetosheath, but still far from a realistic SW size, which is currently impossible to reproduce with kinetic codes. The chosen parameters are designed to represent typical SW conditions \citep{bandyopadhyay2020situ}. The simulation time step is $dt\!=\!0.01\,\Omega^{-1}_i$ (with $\Omega_i$ the ion cyclotron frequency). We employ a numerical filter to smooth electromagnetic fluctuations on scales of order the grid step size, improving numerical stability. Additional information are provided in \citet{arro2024large}.

\section{Results}

\subsection{Secondary Electron-Only reconnection Produced by Large Scale Ion-Coupled Reconnection}
\label{sec:Secondary_EREC.}

\begin{figure*}[bpt]
\centering
\includegraphics[width=\linewidth]{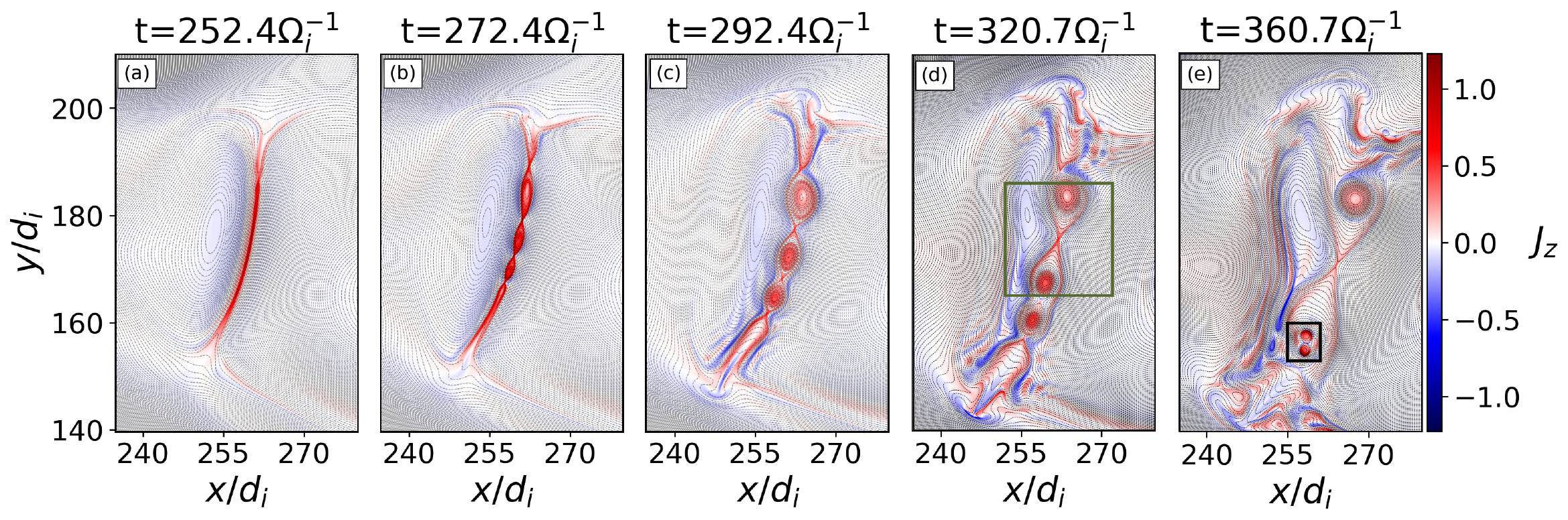}
\caption{Panels (a)-(e): shaded isocontours of the out-of-plane current density $J_z$ (normalized to $J_0 = e n_0 c_{A,0}$) over a sub-region of the simulation box at times $t\,\Omega_i \in [252.4,\,272.4,\,292.4,\,320.7,\,360.7]$, with black dashed lines representing in-plane magnetic field lines. The green and black squares in panels (d) and (e) indicate the regions where IREC and EREC take place, respectively.}
\label{fig:FIG1_tearing_instability_evolution}
\end{figure*}

In this subsection, we analyze the development of a secondary EREC event within the outflow of a large-scale IREC site in our simulation of SW turbulence. We highlight and discuss the properties and signatures of both reconnection regimes, which behave differently due to the different scales involved.

Figure \ref{fig:FIG1_tearing_instability_evolution} illustrates the temporal evolution of the out-of-plane current density, $J_z$, in a sub-region of the simulation box. $J_z$ is shown as shaded isocontours and is normalized to $J_0 = e n_0 c_{A,0}$, where $e$ is the proton charge and $c_{A,0}=B_0/\sqrt{4\pi m_i n_0}$ is the initial Alfvén speed, while black dashed lines indicate the in-plane magnetic field. This normalization for $J_z$ will be used throughout the whole paper. The different panels progressively follow the evolution of a CS as it undergoes reconnection. This CS, whose initial length is between $40\,d_i$ and $50\,d_i$, as seen in Fig. \ref{fig:FIG1_tearing_instability_evolution}(a), forms spontaneously because of the large-scale turbulent dynamics. By $t=272.4\,\Omega_i^{-1}$, Fig. \ref{fig:FIG1_tearing_instability_evolution}(b), the CS becomes unstable and reconnects, forming three plasmoids. It is well-known that CSs are prone to the linear tearing instability \citep{Furth_etal_1963,Bulanov_etal_1992,Chen_etal_1997,Uzdensky_etal_2016}, leading to reconnection, with the consequent formation of plasmoid chains \citep{loureiro2007instability,uzdensky2010fast}, playing a significant role in plasma turbulence \citep{Walker_etal_2018,Boldyrev_etal_2018}. As the instability proceeds and enters the non-linear stage, plasmoids grow bigger and saturate, as seen in Fig. \ref{fig:FIG1_tearing_instability_evolution}(c), at $t=292.4\Omega_i^{-1}$, where plasmoids have reached a size ranging between $5\,d_i$ and $10\,d_i$. Panel (d), at $t=320.7\,\Omega_i^{-1}$, shows that a main X-point has developed between the two upper plasmoids, where IREC occurs, as we will show later. The lower plasmoids, situated in the IREC outflow, are pushed one against the other by the reconnection jet. A new reconnecting CS with length smaller than $2d_i$ forms between two lower plasmoids, as seen in Fig. \ref{fig:FIG1_tearing_instability_evolution}(e) and by the black box, at $t=360.7\,\Omega_i^{-1}$. As we will show later, this secondary CS hosts an EREC event, whose outflow is orthogonal to the initial magnetic shear of the primary large-scale IREC CS.

\begin{figure*}[tbp]
\centering
\includegraphics[width=\linewidth]{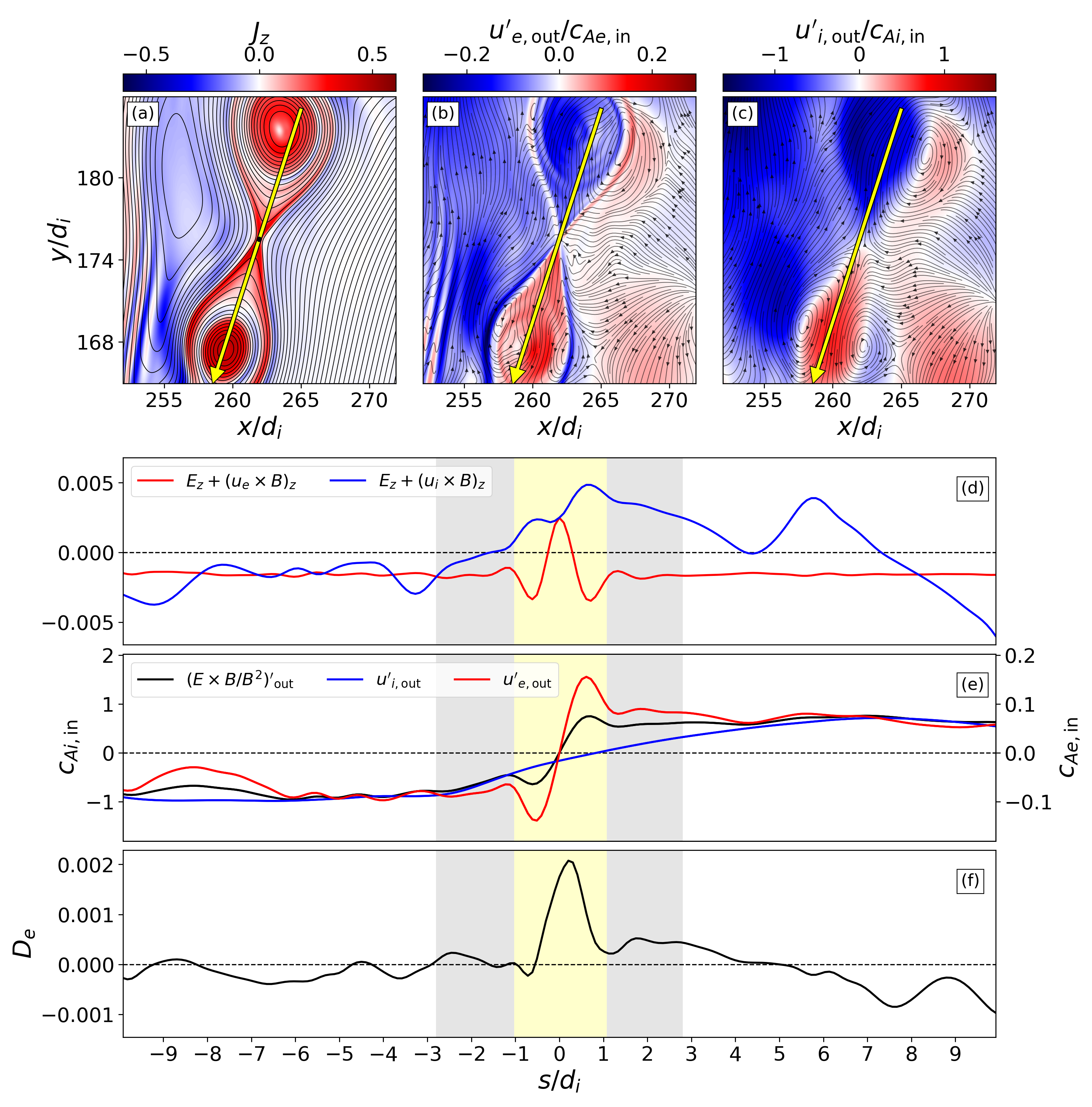}
\caption{Analysis of the large-scale reconnection region at $t=320.7\Omega_i^{-1}$ (green square in Fig. \ref{fig:FIG1_tearing_instability_evolution}d). 
Panels (a)-(c): Shaded isocontours where the yellow arrow indicates the direction of the 1D cut along the exhaust. 
(a) Out-of-plane current density $J_z$ (normalized to $J_0$) with in-plane magnetic field lines (black dashed lines). 
(b)-(c) Outflow-directed electron ($u'_{e,\text{out}}$) and ion ($u'_{i,\text{out}}$) fluid velocities in the X-point frame. 
Panels (d)-(f): Profiles along the 1D cut defined by the yellow arrow. 
(d) Out-of-plane electric field in the electron frame ($R_{ez} = E_z + (\mathbf{u}_e \times \mathbf{B})_z$, red) and ion frame ($R_{iz} = E_z + (\mathbf{u}_i \times \mathbf{B})_z$, blue), in units of $B_0 c_{A0}/c$. 
(e) Outflow-directed velocities for electrons ($u'_{e,\text{out}}$, red), ions ($u'_{i,\text{out}}$, blue), and the $\mathbf{E}\times \mathbf{B}$ drift (black) in the X-point frame. 
(f) Electron-frame dissipation measure $D_e = \mathbf{J}\cdot \mathbf{R}_e = \mathbf{J}\cdot \left(\mathbf{E}+\mathbf{u}_e\times\mathbf{B}\right)$, normalized to $c_{A,0}J_0B_0/c$. 
The yellow and gray shaded areas highlight the EDR and IDR, respectively.}
\label{fig:FIG2_IREC}
\end{figure*}

\begin{figure*}[tbp]
\centering
\includegraphics[width=\linewidth]{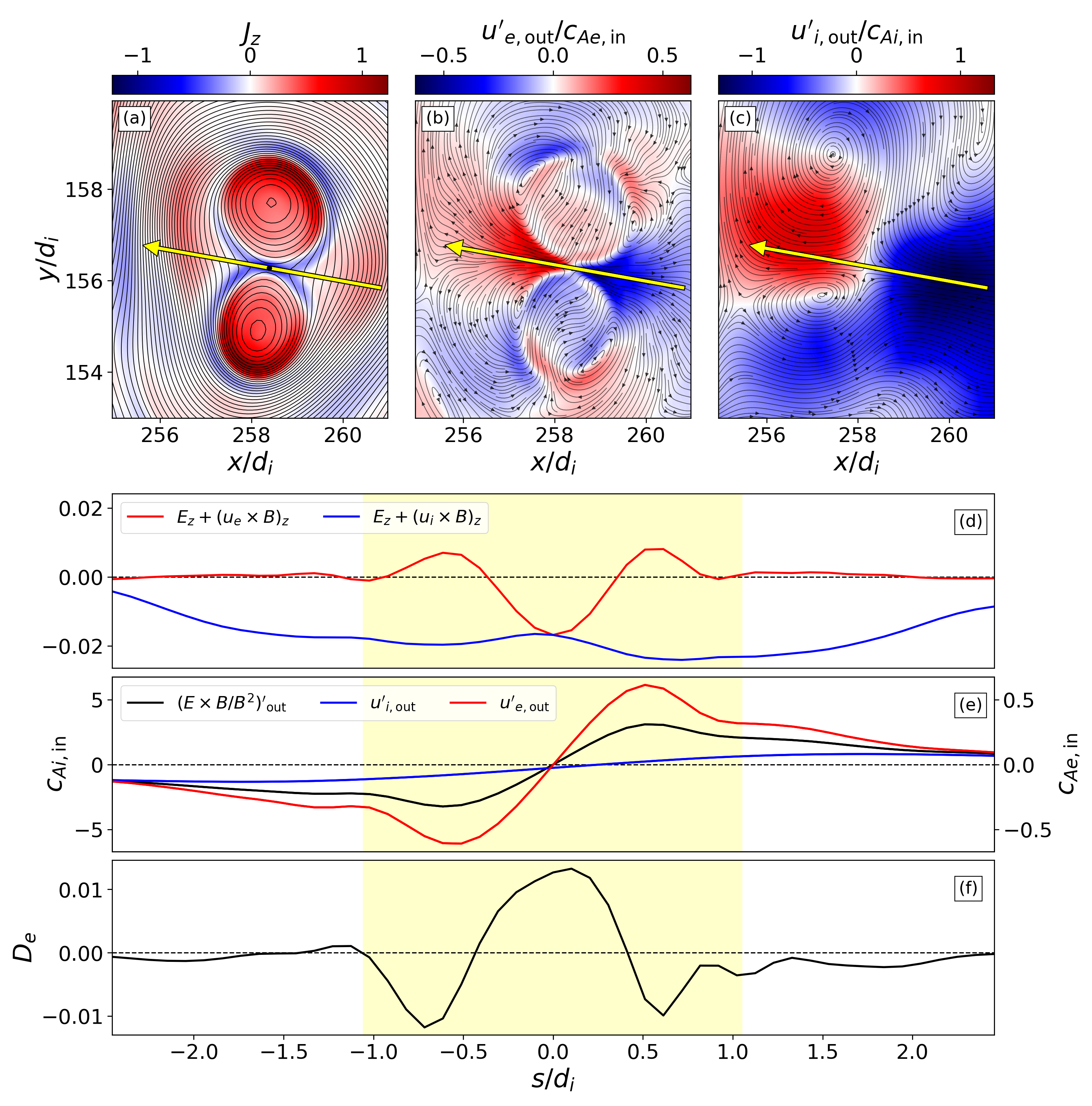}
\caption{Analysis of the small-scale reconnection region at $t=360.7\Omega_i^{-1}$ (black square in Fig. \ref{fig:FIG1_tearing_instability_evolution}e). 
Panels (a)-(c): Shaded isocontours where the yellow arrow indicates the direction of the 1D cut along the exhaust. 
(a) Out-of-plane current density $J_z$ (normalized to $J_0$) with in-plane magnetic field lines (black dashed lines). 
(b)-(c) Outflow-directed electron ($u'_{e,\text{out}}$) and ion ($u'_{i,\text{out}}$) fluid velocities in the X-point frame. 
Panels (d)-(f): Profiles along the 1D cut defined by the yellow arrow. 
(d) Out-of-plane electric field in the electron frame ($R_{ez} = E_z + (\mathbf{u}_e \times \mathbf{B})_z$, red) and ion frame ($R_{iz} = E_z + (\mathbf{u}_i \times \mathbf{B})_z$, blue), in units of $B_0 c_{A0}/c$. 
(e) Outflow-directed velocities for electrons ($u'_{e,\text{out}}$, red), ions ($u'_{i,\text{out}}$, blue), and the $\mathbf{E}\times \mathbf{B}$ drift (black) in the X-point frame. 
(f) Electron-frame dissipation measure $D_e = \mathbf{J}\cdot \mathbf{R}_e = \mathbf{J}\cdot \left(\mathbf{E}+\mathbf{u}_e\times\mathbf{B}\right)$, normalized to $c_{A,0}J_0B_0/c$. 
The yellow shaded area highlights the EDR.}\label{fig:FIG3_EREC}
\end{figure*}

We now analyze and characterize the two IREC and EREC sites identified above, highlighted in Figures \ref{fig:FIG1_tearing_instability_evolution}(d) and \ref{fig:FIG1_tearing_instability_evolution}(e), marked by the green and black squares enclosing them. Hereafter, we refer to the green square ($20\,d_i\times 21\,d_i$) as the \enquote{large-scale reconnection region} and the black square ($6\,d_i\times 7\,d_i$) as the \enquote{small-scale reconnection region}. Panels (a) of Figures \ref{fig:FIG2_IREC} and \ref{fig:FIG3_EREC} show $J_z$ isocontours for the large-scale and small-scale regions, respectively. Black lines represent the in-plane magnetic field, and a black dot denotes the X-point. The 1D cuts, indicated by yellow arrows in panels (a)-(c) of both figures, are aligned to the outflow of their respective reconnection sites. This outflow direction is defined as the eigenvector associated with the minimum eigenvalue of the magnetic-flux function's Hessian matrix at the X-point \citep{servidio2009magnetic}. 

The distinction between IREC and EREC is often related to the scale of the reconnecting magnetic structures. \citet{Pyakurel_etal_2019} studied this transition in laminar reconnection by varying the simulation domain size (which is considered equivalent to the size of magnetic islands in turbulence) and showed a gradual transition from EREC to IREC as the size increased from $\sim 4d_i$ to $\sim 40d_i$. According to their study, the domain size is a critical factor determining whether IREC or EREC occurs, because achieving a fully frozen-in ion exhaust, with ion flows comparable to the Alfvén speed, requires an exhaust width of at least several $d_i$. 

In our turbulence simulation, we estimate the maximum width of the exhaust along the direction normal to the CS to be the width of the plasmoids involved in the reconnection process. For the large-scale region, the exhaust width is $\sim 5$-$10d_i$ (Fig. \ref{fig:FIG2_IREC}a), while for the small-scale region, it is less than $\sim 2d_i$ (Fig. \ref{fig:FIG3_EREC}a). \citet{Pyakurel_etal_2019} found that the smallest discernible ion jet requires an exhaust width of at least $\sim 2d_i$, and fully ion-coupled reconnection requires $\sim 8d_i$. Consistent with this picture, we expect the large-scale region to couple with ions, driving IREC. In contrast, the small-scale region should exhibit EREC, as ions are likely decoupled from the small-scale dynamics of electrons and the magnetic fields. To verify this hypothesis, we first analyze the large-scale reconnection region and then compare it to the small-scale reconnection region.

To identify reconnection signatures, we use 1D cuts (indicated by yellow arrows). Panels (d) of Figures \ref{fig:FIG2_IREC} and \ref{fig:FIG3_EREC} show the out-of-plane electric field in the frame of the ions and electrons, respectively: $\mathbf{R}_i = \mathbf{E}+(\mathbf{u}_i\times \mathbf{B})$ (blue) and $\mathbf{R}_e = \mathbf{E}+(\mathbf{u}_e\times \mathbf{B})$ (red), in units of $B_0 c_{A0}/c$ (with $\textbf{u}_i$, with $\textbf{u}_e$ being the ion and electron fluid velocities). Breaking the magnetic flux frozen-in condition for a species $\alpha$ (to allow the disruption of the magnetic field lines topology) requires $\nabla\times \mathbf{R}_\alpha \neq 0$ \citep{Goldman_etal_2016}. In our 2D case (where $\partial/\partial z = 0$), only the out-of-plane component $R_{\alpha z}$ is relevant for breaking the in-plane flux conservation \citep{Califano_etal_2020}. Localized regions in both $R_{iz}$ and $R_{ez}$ are apparent in the structure and, thus, the Electron Diffusion Region (EDR) and Ion Diffusion Region (IDR) are found where $\nabla R_{ez} \neq 0$ and $\nabla R_{iz}\neq 0$, respectively. 

\subsubsection{Ion coupled reconnection site}

We first analyze the large-scale reconnection region (Fig. \ref{fig:FIG2_IREC}). Panel (b) shows the outflow-directed components of the electron ($u'_{e,\text{out}}$) and ion ($u'_{i,\text{out}}$) fluid velocities, respectively, in the frame of the X-point. The yellow arrow indicates the direction of the cut. These velocities are normalized to the inflow Alfvén velocity $c_{Ai,\text{in}}=B_{\text{in}}/\sqrt{4\pi m_i n_{\text{in}}}$ (details in Appendix A). The black streamlines in panels (b) and (c) represent the in-plane electron and ion fluid velocities in the X-point frame. The X-point velocity itself, $\mathbf{u}_X$, is the $\textbf{E}\times\textbf{B}$ drift velocity evaluated at the X-point, i.e. $\mathbf{u}_\text{X} = \left.\mathbf{E}\times\mathbf{B}/|\mathbf{B}|^2\right|_{\text{X-point}}$. Thus, the velocities in the X-point frame are $\mathbf{u}'_{e} = \mathbf{u}_e - \mathbf{u}_{\text{X}}$ and $\mathbf{u}'_{i} = \mathbf{u}_i - \mathbf{u}_{\text{X}}$ (Hereafter, $'$ denotes the X-point frame and the subscript "$\text{out}$" denotes the outflow-directed component).

In the large-scale reconnection site, clear ion jets flow out of the reconnection region, strongly correlated with the magnetic field geometry (Fig. \ref{fig:FIG2_IREC}c). These ion outflows are accompanied by super-Alfvénic electron jets flowing out of the X-point (Fig. \ref{fig:FIG2_IREC}b). These are classic signatures of IREC \citep{Lee_etal_2020}, implying that ions become magnetized in the outflows and are expelled from the reconnection region together with the reconnected magnetic field lines, following the electrons, which are magnetized and expelled earlier.

In the large-scale reconnection case (Fig. \ref{fig:FIG2_IREC}d), $R_{ez}$ is non-constant only in a narrow region of width $\sim 2d_i$ around the X-point, the EDR (yellow shaded area). Outside the EDR, $R_{ez}$ is non-zero but constant, meaning that the magnetic flux is frozen to the electrons, even if the electron slippage condition ($R_{ez} = 0$) is not satisfied \citep{Schindler_etal_1991, Goldman_etal_2016}. $R_{iz}$ is enhanced near the X-point (where ions are demagnetized) but decays to zero far away (where ions are frozen-in). This behavior defines the IDR (gray shaded area), which has a width of $\sim 6d_i$.

Panel (e) of Fig. \ref{fig:FIG2_IREC} shows the outflow-directed velocities $u'_{e,\text{out}}$ (red), $u'_{i,\text{out}}$ (blue), and the $\mathbf{E}\times\mathbf{B}$ drift (black) in the X-point frame along the 1D cut. Velocities are in units of both the inflow ion and electron Alfvén speeds $c_{Ai,\text{in}}$ and $c_{Ae,\text{in}} = (m_e/m_i)^{1/2} c_{Ai,\text{in}}$. The large-scale reconnection site shows super-Alfvénic electron jets near the X-point ($\sim 2c_{Ai,\text{in}}$) in conjunction with the standard Alfvénic ion jets \citep{Lee_etal_2020}. Electron jets slow down to match the $\mathbf{E}\times\mathbf{B}$ drift velocity, defining an exhaust length of $\sim 9d_i$. Notably, electron jets have velocities below the electron Alfvén speed. This is consistent with recent EREC studies \citep{Guan_etal_2024,Liu_etal_2025}, showing that charge separation between demagnetized ions and accelerated electrons creates a Hall electric field that decelerates the electrons, preventing them from reaching $c_{Ae,\text{in}}$, and keeping the reconnection rate $cE_z < 0.1 B_{\text{in}}c_{Ae,\text{in}}$.

Panel (f) of Fig. \ref{fig:FIG2_IREC} shows the electron-frame dissipation measure $D_e = \mathbf{J}\cdot \mathbf{R}_e = \mathbf{J}\cdot \left(\mathbf{E}+\mathbf{u}_e\times\mathbf{B}\right)$ (the \enquote{Zenitani measure} \citep{zenitani2011new,Yang_etal_2024}) along the 1D cut, in units of $c_{A0}J_0B_0/c$. $D_e$ quantifies the rate of magnetic-to-plasma energy conversion in the electron frame. The large-scale reconnection region shows enhanced energy conversion centered at the X-point, as expected for standard IREC.

\subsubsection{Electron-only reconnection site}

We now proceed to analyze the small-scale reconnection region in comparison with the standard IREC case. Figure \ref{fig:FIG3_EREC} presents this analysis using the same format as Fig. \ref{fig:FIG2_IREC}. Similarly to the large-scale reconnection case, the small-scale site exhibits a super-Alfvénic electron jet flowing out of the X-point (Fig. \ref{fig:FIG3_EREC}b). However, the ion dynamics differ substantially. In contrast to the large-scale region, the small-scale reconnection site shows no ion jets correlated to the electron outflow (Fig. \ref{fig:FIG3_EREC}c), and ion streamlines (in black) follow a flow pattern which is not correlated to magnetic field lines. With an exhaust width smaller than $2d_i$, reconnection occurs in a region too small for ions to be magnetized and coupled to the reconnection process. As shown in the 1D cut of Fig. \ref{fig:FIG3_EREC}(d), $R_{iz}$ is not constant along the entire exhaust, indicating that ions are never coupled to the magnetic field in this small-scale reconnection event, which is a clear signature of EREC \citep{Guan_etal_2024}. 

Figure \ref{fig:FIG3_EREC}e shows that, like in the large-scale reconnection case, electron jets are super-Alfvénic near the X-point (with maximum speed $\sim 5c_{Ai,\text{in}}$), but do not exceed the electron Alfvén speed due to the decelerating Hall electric field. However, the standard Alfvénic ion jets are absent. Regarding dissipation (Fig. \ref{fig:FIG3_EREC}f), the small-scale region shows a positive peak at the X-point, flanked by two symmetric negative minima (region of plasma-to-magnetic energy transfer). The peak dissipation rate here ($D_e \sim 0.01$) is about one order of magnitude higher than in the large-scale region ($D_e \sim 0.001$). Because ions are decoupled in EREC, magnetic flux can flow more rapidly, leading to a higher reconnection rate and significant dissipation near the X-point. 

The analysis of the properties of reconnection sites indicates that the large- and small-scale reconnection regions are consistent with IREC and EREC, respectively. We have shown that cross-scale interactions can produce secondary EREC in the IREC outflow. Therefore, the process we have described represents a possible mechanism for the spontaneous formation of EREC in the turbulent SW, without the need for an external small-scale energy injection or a short correlation length (in contrast to what happens in the Earth's magnetosheath).

\subsection{Electron-Only reconnection Produced by the Electron Kelvin-Helmholtz instability}
\label{sec:EKHI_EREC}

\begin{figure*}[tbp]
\centering
\includegraphics[width=\linewidth]{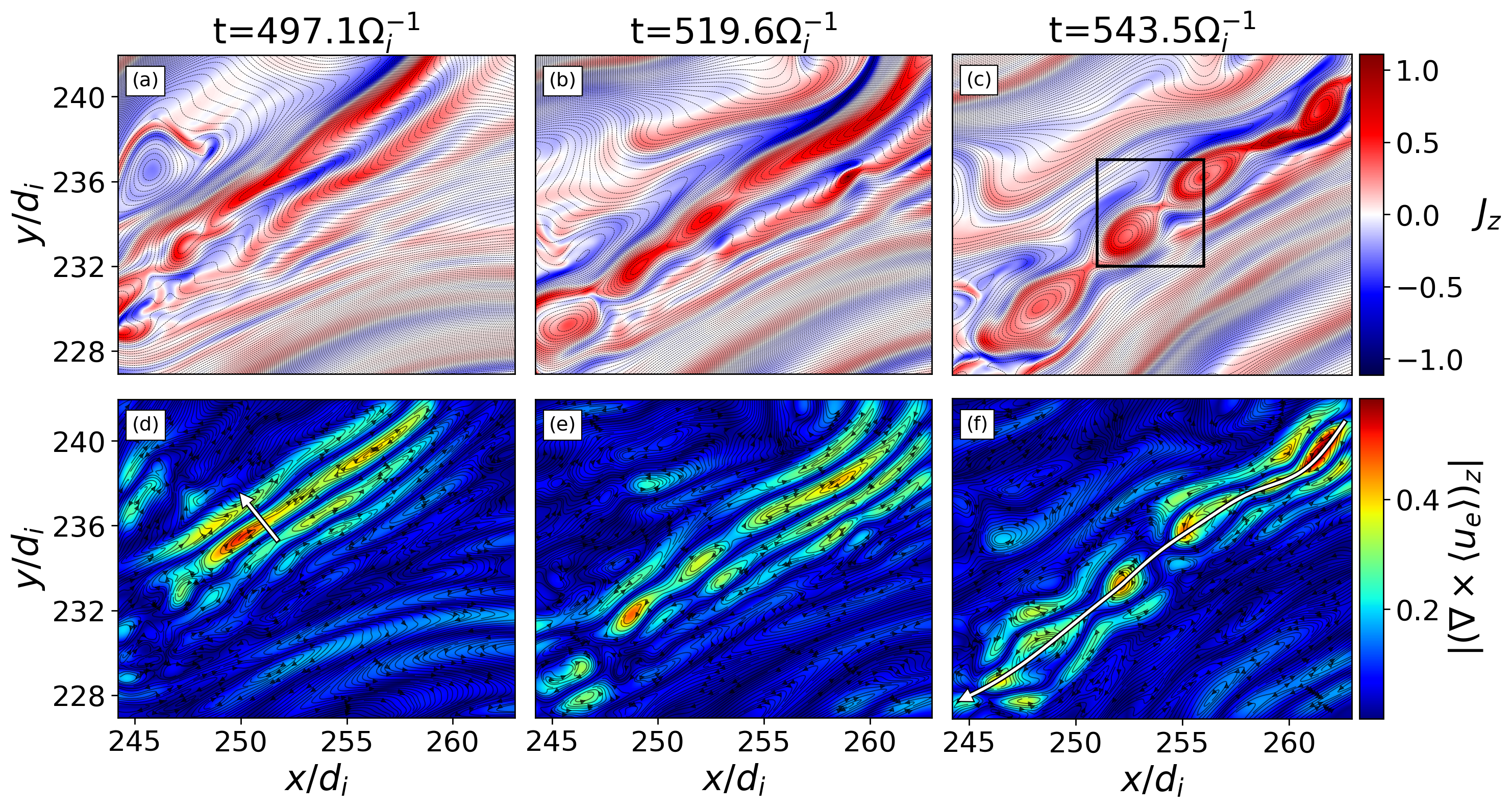}
\caption{Temporal evolution of the EKHI-induced EREC events for times $t=497.1\Omega_i^{-1}$ (first column), $t=519.6\Omega_i^{-1}$ (second column) and $t=543.5\Omega_i^{-1}$ (third column). Panels (a)-(c): Shaded isocontours of the out-of-plane current density $J_z$ (normalized to $J_0$) with in-plane magnetic field lines (black shaded). Panels (d)-(f): Shaded isocontours of the out-of-plane low-pass-filtered electron vorticity magnitude $|(\nabla \times \langle \mathbf{u}_e\rangle)_z|$ (normalized to $c_{A0}/d_i$), with black streamlines representing the in-plane electron velocity $\mathbf{u}_e$ in the frame of the bulk plasma velocity $\mathbf{u}=(m_e \mathbf{u}_e + m_i\mathbf{u}_i)/(m_e+m_i)$.}
\label{fig:FIG4_EKHI}
\end{figure*}

\begin{figure*}[tbp]
\centering
\includegraphics[width=\linewidth]{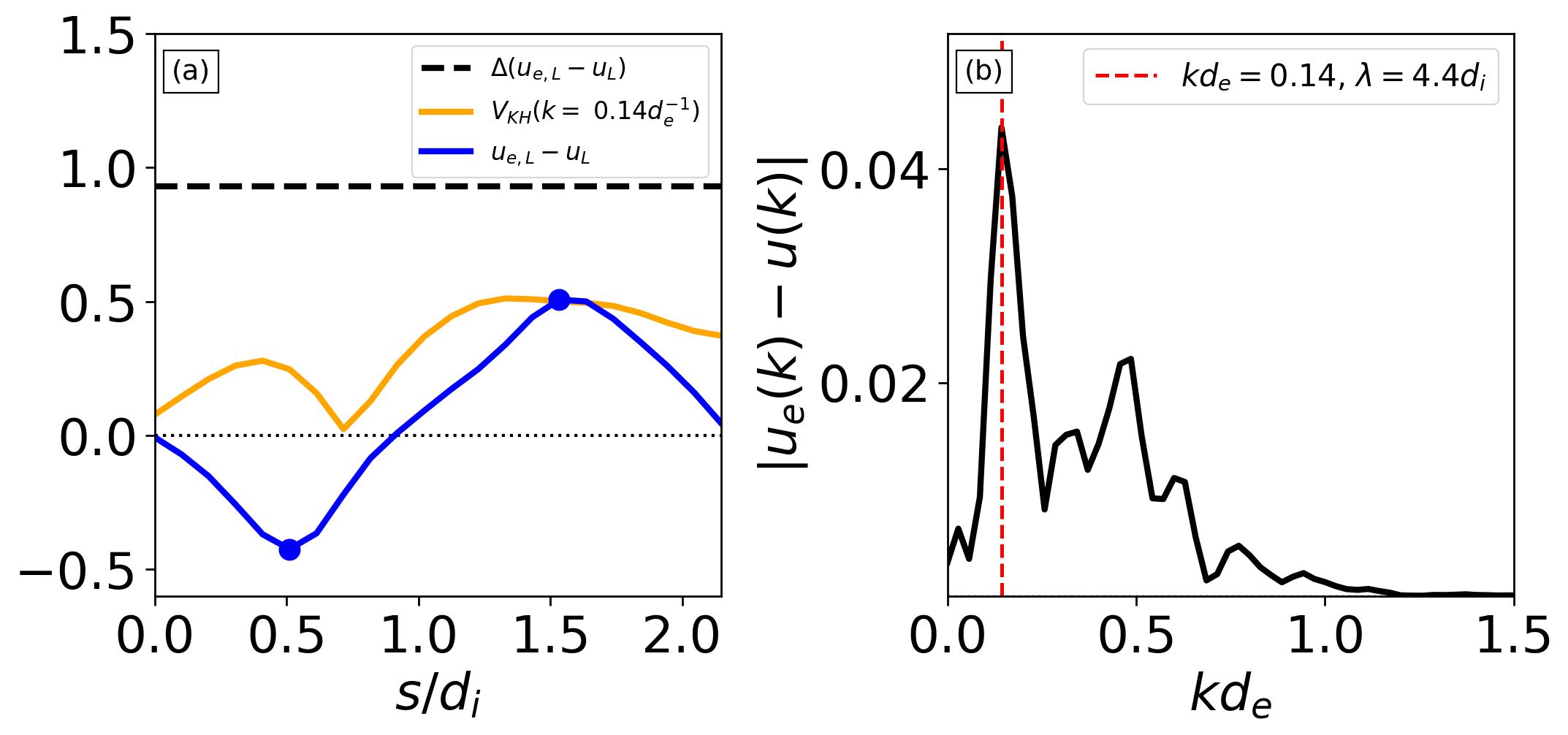}
\caption{(a) Shear-aligned electron velocity in the frame of the bulk velocity $u_{e,L}-u_L$ (blue, normalized to $c_{A0}$), the local threshold velocity $V_{\text{KH}}$ (orange) evaluated at $k = 0.14d_e^{-1}$ and the velocity jump $\Delta(u_{e,L}-u_L)$ (black dashed line). (b) FFT of $u_{e,L}-u_L$ across the white arrow of Fig. \ref{fig:FIG4_EKHI}f, which crosses the vortex chain. The red dashed horizontal line passes through the peak wavenumber at $kd_e \sim 0.14$.}
\label{fig:FIG5_EKHI_threeshold}
\end{figure*}

In this subsection, we discuss another mechanism for the spontaneous formation of EREC sites in our large-scale turbulence simulation. Unlike the process described in the previous section, this mechanism does not require a precursor large-scale IREC site; EREC develops directly due to the dynamics of electron-scale shear flows produced by turbulent dynamics.

Figure \ref{fig:FIG4_EKHI} illustrates the temporal evolution of a sub-region of the simulation box. The first row shows $J_z$ (shaded isocontours) with in-plane magnetic field lines (black dashed). The second row shows the out-of-plane low-pass-filtered electron vorticity magnitude, $|(\nabla \times \langle \mathbf{u}_e\rangle)_z|$, normalized to $c_{A0}/d_i$. Here, $\langle \rangle$ denotes a spatial Gaussian low-pass filter with a width of $d_i$, chosen to highlight electron vortices while suppressing smaller-scale in-plane currents associated with reconnection. Black streamlines in the second row represent $\mathbf{u}_e$ in the frame of the bulk plasma velocity $\mathbf{u}  = (m_e \mathbf{u}_{e}+m_i \mathbf{u}_{i})/(m_e+m_i)$. The figure shows the formation of a plasmoid chain in which magnetic field lines wrap around near-ion-scale electron vortices. In the following, we show that interactions among these plasmoids produce EREC. 

At $t=497.1\Omega_i^{-1}$ (early stage), plasmoids have not yet formed, but magnetic field lines are beginning to bend (Fig. \ref{fig:FIG4_EKHI}a). This region exhibits strong electron velocity shears, evidenced by enhanced electron vorticity (Fig. \ref{fig:FIG4_EKHI}d). By $t=519.6\Omega_i^{-1}$, electron vortices begin to emerge (Fig. \ref{fig:FIG4_EKHI}e). These structures are near-ion-scale (i.e., with wavelength $\lambda \sim d_i$), a regime well-described by electron magnetohydrodynamics (EMHD) \citep{Bulanov_etal_1992}. In this regime, ions are demagnetized and follow separate dynamics from the electrons, which remain coupled to the magnetic field. Consequently, the frozen-in magnetic flux wraps around the electron vortices, driving the formation of plasmoids (Fig. \ref{fig:FIG4_EKHI}b). By $t=543.5\Omega_i^{-1}$, plasmoids are fully formed (Fig. \ref{fig:FIG4_EKHI}c), creating a chain of reconnection X-points. Notably, each plasmoid is embedded within an electron vortex (Fig. \ref{fig:FIG4_EKHI}f). 

The disruption of the velocity shear into electron vortices is qualitatively consistent with the development of the Electron Kelvin-Helmholtz Instability (EKHI). This instability has been studied extensively in cold EMHD \citep{Das_etal_2003,Jain_etal_2003,Jain_etal_2004,Gaur_etal_2009,Gaur_etal_2012} and in fully kinetic simulations and satellite observations of magnetic reconnection \citep{Pritchett_etal_2009,Fermo_etal_2012,Huang_etal_2015,Zhong_etal_2018,Zhong_etal_2022}, where it disrupts electron velocity shears in diffusion regions and secondary reconnection outflows. The EKHI has previously been shown to contribute to generating other types of kinetic scale structures in turbulent plasma, such as subion scale magnetic holes \citep{arro2023generation}. Using an EMHD model, \citet{Gaur_etal_2009} demonstrated that whistler waves can suppress the EKHI. Flow perturbations from the EKHI distort the magnetic field into a sheared configuration, exciting whistler oscillations via the restoring magnetic tension force. These oscillations extract energy from the velocity shear, stabilizing the flow. Therefore, for the EKHI to develop in a velocity shear with thickness $l$, the EKHI characteristic growth rate $\gamma \sim \Delta u_{e,L}/l$  must exceed the frequency of whistler waves propagating parallel to the shear flow $\omega = \Omega_{e,L}k^2d_e^2/(1+k^2d_e^2)$\citep{Fermo_etal_2012,arro2023generation}. Here, $\Delta u_{e,L}$ is the jump in the electron velocity across the shear, and $\Omega_{e,L}$ is the electron cyclotron frequency based on the shear-aligned magnetic field.  By imposing that $\gamma$ must be larger than $\omega$ for the EKHI to be able to get excited, the instability criterion is obtained:
\begin{equation}
    \Delta u_{e,L} > V_{\text{KH}}(k) := \left(\frac{l}{d_e}\right)c_{Ae,L}\frac{k^2 d_e^2}{1+k^2d_e^2},
\label{eq:EKHI_threeshold}
\end{equation}
where $c_{Ae,L} = d_e\Omega_{e,L}$ is the shear-aligned electron Alfvén speed, and $k$ is the wavenumber of the unstable mode. For a shear width $l\sim d_e$ we recover the result of \citet{arro2023generation}, and by also considering fluctuations with $k\sim d_e^{-1}$ , Eq. (\ref{eq:EKHI_threeshold}) gives the criterion $\Delta u_{e,L} > c_{Ae,L}/2$ used by \citet{Fermo_etal_2012}.

\begin{figure*}[tbp]
\centering
\includegraphics[width=\linewidth]{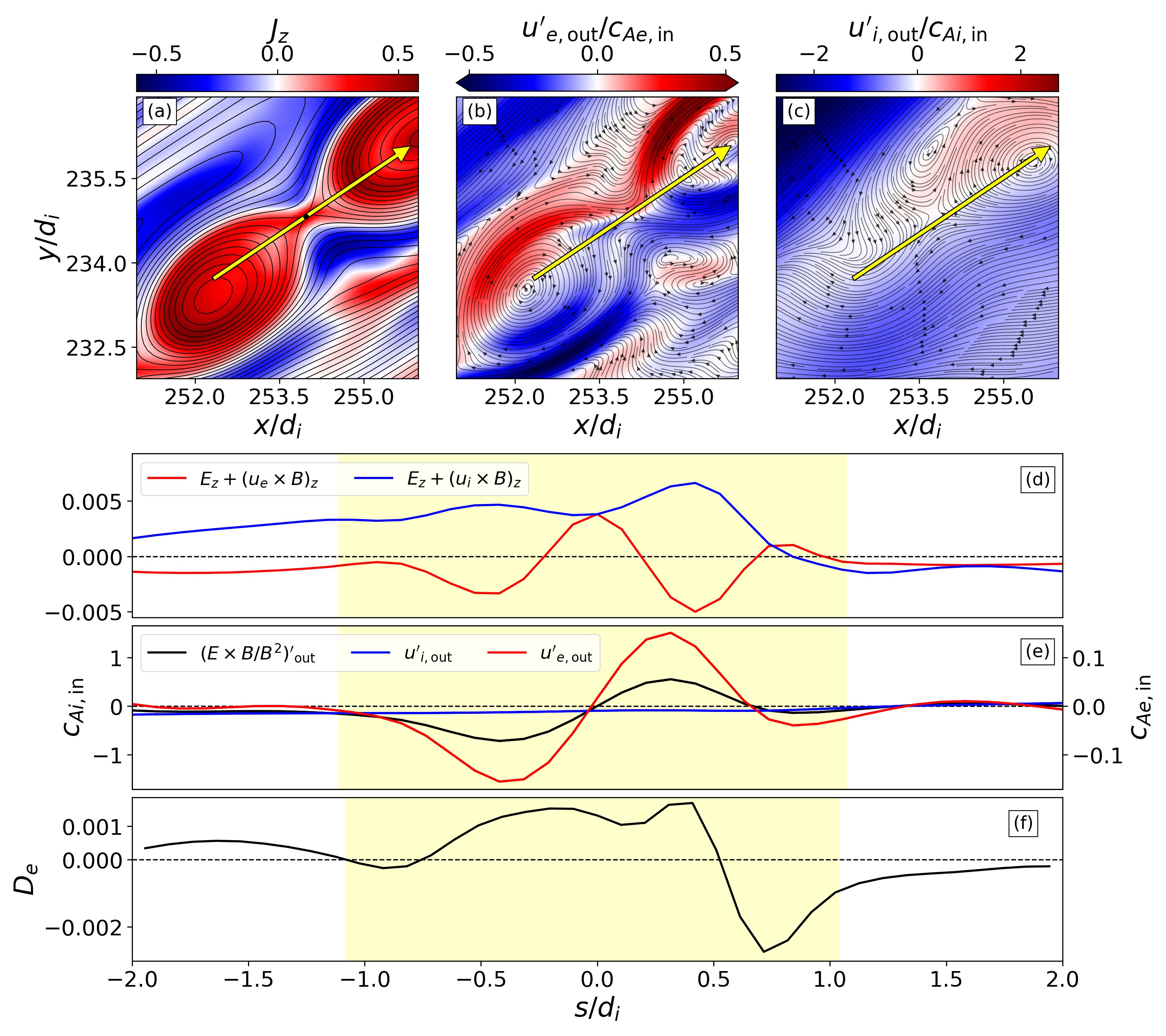}
\caption{Analysis of one of the reconnection sites (black square in Fig. \ref{fig:FIG4_EKHI}c) embedded in the electron vortices at $t=543.5\Omega_i^{-1}$. 
Panels (a)-(c): Shaded isocontours where the yellow arrow indicates the direction of the 1D cut along the exhaust. 
(a) Out-of-plane current density $J_z$ (normalized to $J_0$) with in-plane magnetic field lines (black dashed). 
(b)-(c) Outflow-directed electron ($u'_{e,\text{out}}$) and ion ($u'_{i,\text{out}}$) fluid velocities in the X-point frame. 
Panels (d)-(f): Profiles along the 1D cut defined by the yellow arrow. 
(d) Out-of-plane electric field in the electron frame ($R_{ez} = E_z + (\mathbf{u}_e \times \mathbf{B})_z$, red) and ion frame ($R_{iz} = E_z + (\mathbf{u}_i \times \mathbf{B})_z$, blue), in units of $B_0 c_{A0}/c$. 
(e) Outflow-directed velocities for electrons ($u'_{e,\text{out}}$, red), ions ($u'_{i,\text{out}}$, blue), and the $\mathbf{E}\times \mathbf{B}$ drift (black) in the X-point frame. 
(f) Electron-frame dissipation measure $D_e = \mathbf{J}\cdot \mathbf{R}_e = \mathbf{J}\cdot \left(\mathbf{E}+\mathbf{u}_e\times\mathbf{B}\right)$, normalized to $c_{A0}J_0 B_0/c$. 
The yellow shaded area highlights the EDR.}
\label{fig:FIG6_EREC}
\end{figure*}

To confirm that the observed instability is quantitatively consistent with the EKHI, we perform a local stability analysis immediately prior to the formation of electron vortices. For this, we perform a transversal 1D cut through the electron velocity shear region (at $t=497.1$, white arrow in figure \ref{fig:FIG4_EKHI}d). Figure \ref{fig:FIG5_EKHI_threeshold}a shows the shear-aligned electron velocity $u_{e,L}$ (blue, normalized to $c_{A0}$) in the frame of the bulk velocity $u_L$, and the threshold velocity $V_{\text{KH}}$ (orange) from Eq. (\ref{eq:EKHI_threeshold}). The shear-aligned direction is defined as $\hat{\mathbf{l}} = \hat{\mathbf{n}}\times\hat{\mathbf{z}}$, where $\hat{\mathbf{n}}$ is the direction of the 1D cut (white arrow in Fig. \ref{fig:FIG4_EKHI}d). The velocity jump, $\Delta (u_{e,L}-u_L)\sim c_{A0}$ (indicated by the dashed line), and the shear thickness, $l\sim d_i$, are calculated as the distance between the points of maximum and minimum velocity across the shear, i.e., the distance between the highlighted blue dots in the figure. Since the plasma is inhomogeneous, $V_{\text{KH}}$ is calculated using the local shear-aligned magnetic field and density. We estimate the wavenumber $k = 2\pi/\lambda$ from the size $\lambda$ of the fully developed vortices. Figure \ref{fig:FIG5_EKHI_threeshold}b shows the fast-Fourier-transform (FFT) of $u_{e,L}-u_L$ across the vortex chain (white arrow in Fig. \ref{fig:FIG4_EKHI}f). The spectrum peaks at $kd_e \sim 0.14$, corresponding to a near-ion-scale vortex size $\lambda \sim 4.4d_i$. Figure \ref{fig:FIG5_EKHI_threeshold}a shows that prior to vortex formation, $\Delta (u_{e,L}-u_L)$ significantly exceeds $V_{\text{KH}}$. This confirms that the region is unstable to the EKHI, supporting our conclusion that turbulence-driven electron velocity shears drive the formation of electron vortices and subsequent reconnection sites. 

Finally, we characterize the nature of these reconnection sites in their developed stage ($t=543.5\Omega_i^{-1}$) by analyzing the region marked by the black square ($5d_i\times5d_i$) in Fig. \ref{fig:FIG4_EKHI}c. Figure \ref{fig:FIG6_EREC} presents this analysis using the same format and normalization as the previous subsection. The plasmoids involved have a width of $\sim d_i$ (Fig. \ref{fig:FIG6_EREC}a); thus, consistently with \citet{Pyakurel_etal_2019}, we expect this region to exhibit EREC. Indeed, we observe super-Alfvénic electron jets (Fig. \ref{fig:FIG6_EREC}b) but no correlated ion jets (Fig. \ref{fig:FIG6_EREC}c), indicating that ions are decoupled from the magnetic field. Although the surrounding vortices strongly distort the electron jets (as shown by the electron streamlines in Fig. \ref{fig:FIG6_EREC}b), the evolution of the magnetic field topology between $t=497.1\Omega_i^{-1}$ and $t=543.5\Omega_i^{-1}$ (Fig. \ref{fig:FIG4_EKHI}a,c) reveals clear signatures of magnetic reconnection.

The 1D cut analysis (yellow arrow) further confirms the EREC nature of the event. The EDR has a width of $\sim 2d_i$ (non-constant $R_{ez}$, Fig. \ref{fig:FIG6_EREC}d), while $R_{iz}$ varies along the entire region, meaning that ions are demagnetized. The electron outflow velocity is $\sim 0.1 c_{Ae,\text{in}}$ (Fig. \ref{fig:FIG6_EREC}e), which is typical for EREC, as discussed previously \citep{Guan_etal_2024,Liu_etal_2025,ren2025anisotropic}. Interestingly, the electron-frame dissipation measure $D_e$ (Fig. \ref{fig:FIG6_EREC}f) is an order of magnitude lower than in the secondary EREC case analyzed in the previous subsection. The lower dissipation is likely due to the reconnection site being embedded within an EKHI-driven vortex. As the Kelvin-Helmholtz mode extracts energy from the shear to drive the instability, less magnetic energy may be available to energize the plasma via reconnection. 

In summary, our analysis confirms that the reconnection sites within the train of near-ion-scale plasmoids (Fig. \ref{fig:FIG4_EKHI}c) are EREC events driven by the EKHI. Our results demonstrate that the EKHI offers an additional pathway through which EREC events can be excited within a large-scale turbulent cascade. We argue that a mixture of secondary instabilities, such as those illustrated in Sections \ref{sec:Secondary_EREC.} and \ref{sec:EKHI_EREC}, can generate EREC events within turbulent plasmas, such as the SW, with correlation lengths much larger than the ion scales. 

\section{Statistical Analysis}

\begin{figure*}[tbp]
\centering
\includegraphics[width=\linewidth]{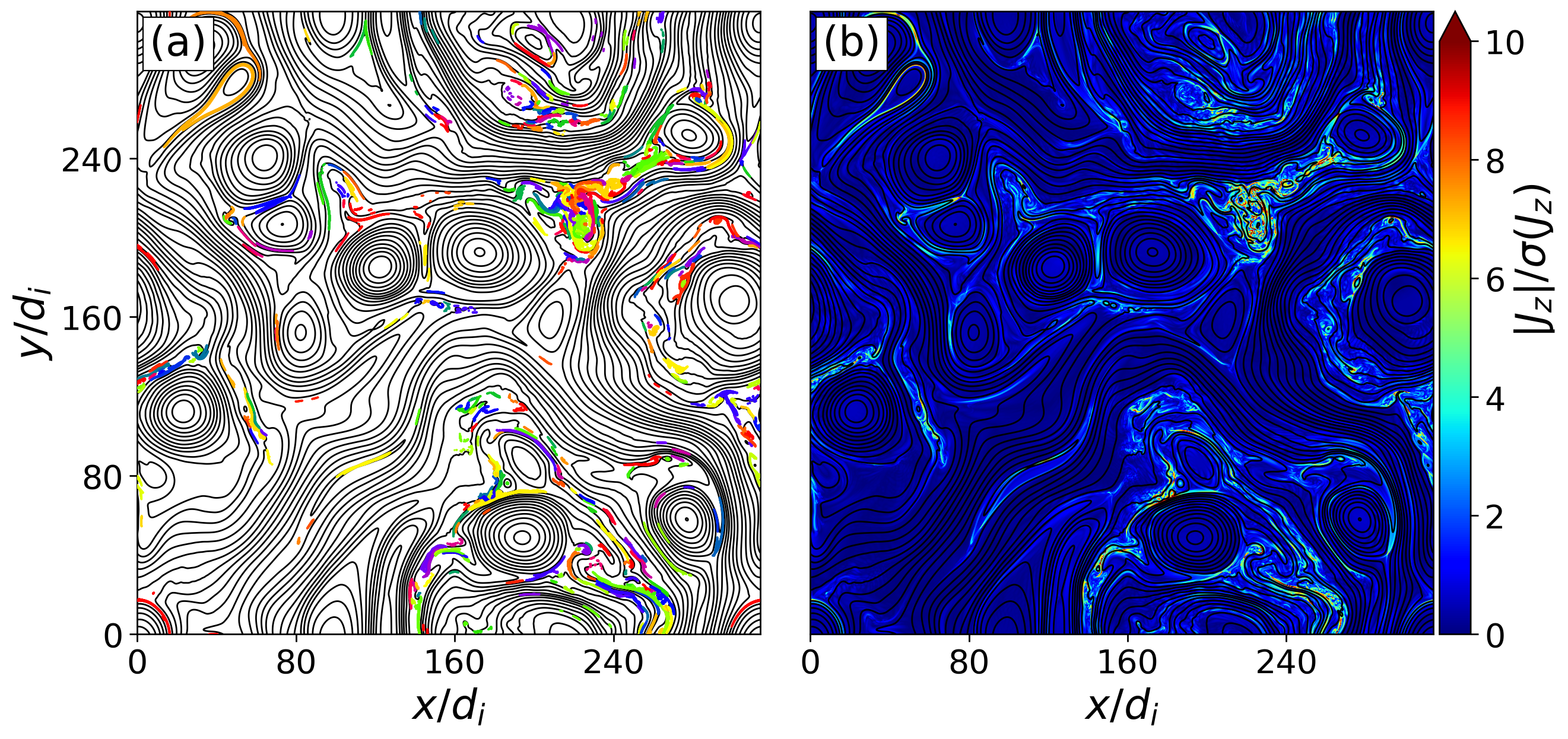}
\caption{Analysis of HDBSCAN clusters at time $t=534.6\Omega_i^{-1}$. Panel (a): In-plane magnetic field lines (black) with identified clusters highlighted in various colors. Panel (b): Shaded isocontours of the normalized out-of-plane current density $|J_z|/\sigma(J_z)$.}
\label{fig:FIG7_Jz_clusters}
\end{figure*}

\begin{figure*}[tbp]
\centering
\includegraphics[width=\linewidth]{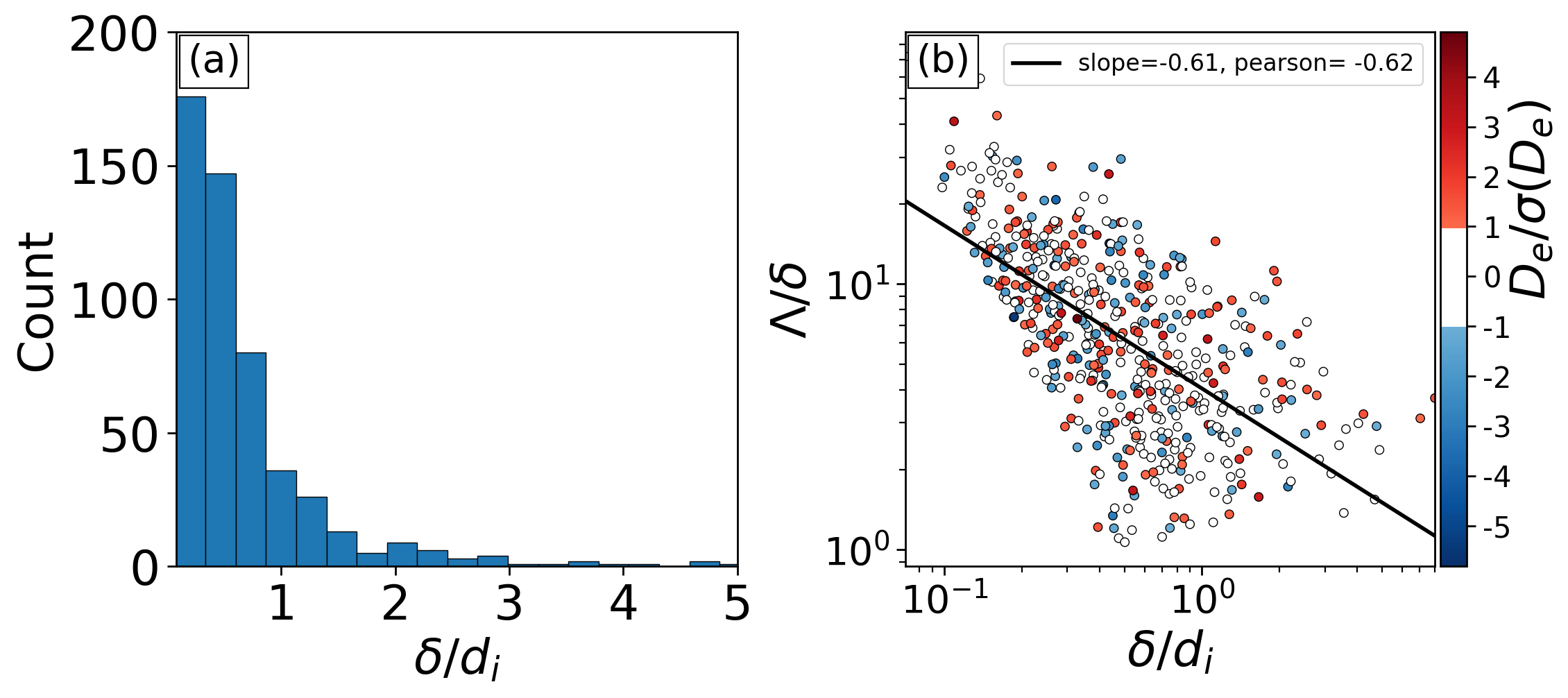}
\caption{Analysis of statistical properties of the HDBSCAN CS clusters at time $t=534.6\Omega_i^{-1}$. Panel (a): Histogram of CSs thicknesses $\delta/d_i$. Panel (b): Scatter plot of the aspect ratio $\Lambda/\delta$ versus $\delta$ (in log-log scale) for each CS, with points colored by their mean $D_e/\sigma(D_e)$. Points with $|D_e/\sigma(D_e)|<1$ are shown in white.}
\label{fig:FIG8_cluster_statistics}
\end{figure*}

In this section, we perform a statistical analysis of the CSs generated by the turbulence. By cataloging these structures from the smallest to the largest scales and identifying their mean properties, we extract statistical information regarding the composition of CSs and their impact on energy conversion. This global perspective complements the local analysis of specific reconnection sites presented in previous sections. To achieve this, we employ HDBSCAN \citep{Campello_etal_2013}, a hierarchical density-based clustering algorithm, to identify individual CSs, following an approach similar to \citet{sisti2021detecting,Lapenta_etal_2022_dbscan}. 

HDBSCAN is an extension of DBSCAN \citep{Ester_etal_1996,Schubert_etal_2017}. While DBSCAN groups tightly packed points in high-density regions based on a fixed distance threshold, HDBSCAN eliminates the need to manually select this parameter. Instead, HDBSCAN generates a cluster hierarchy and selects significant clusters based on their robustness. 

Initializing HDBSCAN requires two parameters: a minimum cluster size and a parameter $p$ that specifies the number of nearest neighbors (including the point itself) to consider when computing the local density of points on the simulation grid. We set the minimum cluster size to the number of grid points contained within a circle of diameter $1\,d_i$; consequently, structures with an area smaller than $\pi d_i^2/4$ are considered \enquote{noise} in the jargon of clustering algorithms, and thus ignored. We choose $p=5$, corresponding to the number of points contained within a circle of radius equal to the physical grid step in our simulation. To select the grid points to be grouped into clusters by HDBSCAN, we apply a threshold to the out-of-plane current density magnitude, selecting points where $|J_z|>3\sigma(J_z)$, where $\sigma(J_z)$ is the standard deviation of $J_z$ over the entire simulation box. Given the 2D geometry considered here, we use $J_z$ because it is the quantity most representative of magnetic reconnection. 

Figure \ref{fig:FIG7_Jz_clusters}a displays the in-plane magnetic field lines at $t=534.6\Omega_i^{-1}$, with identified clusters highlighted in various colors (colors are reused cyclically and serve only to distinguish adjacent clusters). At this time, when turbulence is fully developed, HDBSCAN identifies $516$ clusters, each one corresponding to a CS, ranging in length from a few $d_i$ to several $d_i$. In contrast, at earlier times when the turbulent cascade has not yet reached all scales (e.g., $t=320.7\Omega_i^{-1}$; see Appendix B), fewer and larger CSs are detected ($143$ sheets). Figure \ref{fig:FIG7_Jz_clusters}b shows the shaded isocontours of the normalized out-of-plane current density $|J_z|/\sigma(J_z)$. A comparison between panels (a) and (b) reveals good agreement between the clusters selected by HDBSCAN and the CSs identifiable by eye, validating the algorithm's performance. 

To characterize the morphology of each CS, we define its \enquote{center} $\mathbf{r}_c = (x_c,y_c)$ as the current-weighted centroid $\mathbf{r}_c = \sum_i \mathbf{r}_i |J_z(\mathbf{r}_i)|/\sum_i |J_z(\mathbf{r}_i)|$, where $\mathbf{r}_i = (x_i,y_i)$ are the coordinates of the $i$-th point in the cluster. We define the CS thickness $\delta$ and length $\Lambda$ as twice the square roots of the eigenvalues of the variance matrix
\begin{equation}
M = \frac{\sum_i (\mathbf{r}_i -\mathbf{r}_c) (\mathbf{r}_i -\mathbf{r}_c) |J_z(\mathbf{r}_i)|}{\sum_i |J_z(\mathbf{r}_i)|} ,  
\end{equation}
associating the minimum and maximum eigenvalues to $\delta$ and $\Lambda$, respectively. Figure \ref{fig:FIG8_cluster_statistics}a presents the histogram of CS thicknesses. The distribution is dominated by thin and short structures; specifically, $\sim 69\%$ of CSs exhibit both a thickness $\delta < 1\,d_i$ and a length $\Lambda < 5d_i$ (not shown). Although this method does not distinguish between active reconnection sites and simple CSs, we expect EREC events to be predominantly located within subion scale sheets ($\delta < d_i$).

Figure \ref{fig:FIG8_cluster_statistics}b shows a scatter plot of the aspect ratio $\Lambda/\delta$ versus $\delta$ (log-log scale) for each CS. Each circle in the plot represents a cluster and is colored by the electron-frame dissipation measure averaged over all points in the cluster, $D_e$, normalized to the global standard deviation $\sigma(D_e)$ calculated over the whole simulation domain. Points with $|D_e/\sigma(D_e)|<1$ are shown in white. We observe that as $\delta$ decreases, the aspect ratio increases. The linear regression yields a slope of $-0.61$ with a Pearson correlation coefficient of $-0.62$, indicating a moderate correlation. This trend simply implies that thinner, smaller-scale structures tend to be relatively more elongated. Finally, we find that $\sim 50.2\%$ of CSs exhibit $|D_e/\sigma(D_e)|>1$ (Fig. \ref{fig:FIG8_cluster_statistics}b), suggesting these are highly active structures with intense energy conversion. However, it is important to note that $D_e$ is both positive and negative, resulting in a mean normalized dissipation over all sheets of $\sim 0.02$.

\section{Discussion and conclusions}

In this work, we have investigated the spontaneous formation of EREC using a 2D Hybrid Vlasov-Maxwell simulation of plasma turbulence, initialized with parameters consistent with typical SW conditions. While EREC is commonly observed in Earth's magnetosheath, where the turbulence correlation length is constrained to be near ion scales by bow shock driving mechanisms, its formation in the SW, with a large correlation length, remains an open question. In contrast to previous numerical investigations, our results show that turbulent dynamics can naturally and spontaneously generate EREC events without the need to inject directly at small scales.

We have identified and characterized two distinct mechanisms driving EREC that are associated with the secondary instability of large-scale turbulent structures. The first mechanism involves cross-scale interactions, in which large-scale IREC events produce secondary plasmoids interacting within the IREC outflows, creating small-scale secondary CSs hosting EREC. This process highlights how standard IREC can drive the development of small-scale EREC. In the second mechanism, turbulence drives the formation of subion-scale electron-velocity shears that are unstable to EKHI, leading to electron vortices. These frozen-in electron vortices wrap the magnetic field, creating a chain of near-ion-scale plasmoids that host EREC. This mechanism demonstrates the importance of velocity shears in generating small-scale EREC events within turbulent environments.

The signatures of the EREC sites we have detected are consistent with typical observations, showing super-Alfvénic electron jets and decoupled ions. In particular, the secondary EREC produced by the large-scale IREC event exhibits a magnetic field dissipation rate that is one order of magnitude larger than the dissipation rate induced by the primary IREC, consistent with the faster reconnection rate expected for EREC due to ion demagnetization. In contrast, the EKHI-driven EREC exhibits lower dissipation rates, comparable to those of IREC. This difference suggests that part of the magnetic energy available in the larger-scale magnetic configuration is instead dissipated by the EKHI, resulting in a weaker magnetic-field dissipation rate within the EDR. 

To assess the global relevance of local EREC events, we performed a statistical analysis of CSs using the machine learning tool HDBSCAN, a hierarchical density-based clustering algorithm. Our analysis reveals that the turbulent domain is dominated by thin and short structures: approximately $69\%$ of the identified CSs have thicknesses below the ion inertial length ($\delta < d_i$) and lengths below $5 d_i$, which should be conducive to EREC if they undergo magnetic reconnection. Furthermore, we find that small-scale CSs are characterized by intense local magnetic-to-plasma energy conversion (both positive and negative). This suggests that EREC events could play an important role in dissipating energy at small kinetic scales, even in large-scale systems such as the SW.

This work provides a clear demonstration of several secondary instabilities in 2D that can generate EREC, even within systems with larger magnetic correlation lengths. Effects such as these are expected to operate even in fully 3D systems, and their analysis provides important groundwork for studying the spontaneous generation of EREC in the SW, which captures most of the key physics involved in the process. However, a natural extension to this work will be to investigate the role of EREC in 3D turbulence. In 3D systems, magnetic islands correspond to flux ropes, which can interact in complex ways that are not possible in 2D setups, and additional secondary instabilities may be present that may provide additional avenues for the generation of EREC events or alter the reconnection dynamics. Another extension of our work would be to include electron kinetic effects, which could eventually affect the turbulent cascade at subion scales \citep{matthaeus2020pathways,arro2022spectral}. We plan to extend our results using 2D and 3D fully kinetic simulations in future works.

In conclusion, our findings suggest that EREC is not strictly confined to environments with small magnetic correlation lengths, such as the Earth's magnetosheath. EREC may be a natural and emergent feature of fully developed plasma turbulence even in systems with large correlation lengths, such as the SW. The occurrence of EREC in the SW may be relevant to understanding the nature of non-linear interactions at kinetic scales and the role of reconnection in the dissipation of turbulent fluctuations. Given the current lack of in situ measurements with sufficiently small-scale resolution to unambiguously resolve EREC in the SW, our results demonstrate the necessity of future high-resolution measurements tailored to the solar wind environment in order to resolve essential physics related to the dissipation of turbulence in large astrophysical plasma environments. Ultimately, these results underscore the possible implications of EREC as a collisionless mechanism for local magnetic-to-plasma energy conversion at kinetic scales, highlighting the fundamental role of turbulence-driven cross-scale interactions in the dynamics of the SW and other collisionless space plasmas.

\begin{acknowledgments}
J. E.-T. acknowledges the support of ANID, Chile, through National Doctoral Scholarship No. 21231291. P.S.M. thanks Fondecyt-Chile grant No. 1240281. Numerical simulations and data analysis have been performed on Marconi and Galileo100 at CINECA (Italy), under the ISCRA initiative. Portions of this research were supported by the International Space Science Institute (ISSI) in Bern, through the ISSI International Team project 24-612: Excitation and Dissipation of Kinetic-Scale Fluctuations in Space Plasmas. J.E.S. is supported by the Royal Society University Research Fellowship URF\textbackslash R\textbackslash 251029.
\end{acknowledgments}

\onecolumngrid

\clearpage

\appendix

\section{Calculation of inflow Alfvén velocity and inflow density}

\begin{figure}[htbp]
\centering
\includegraphics[width=\linewidth]{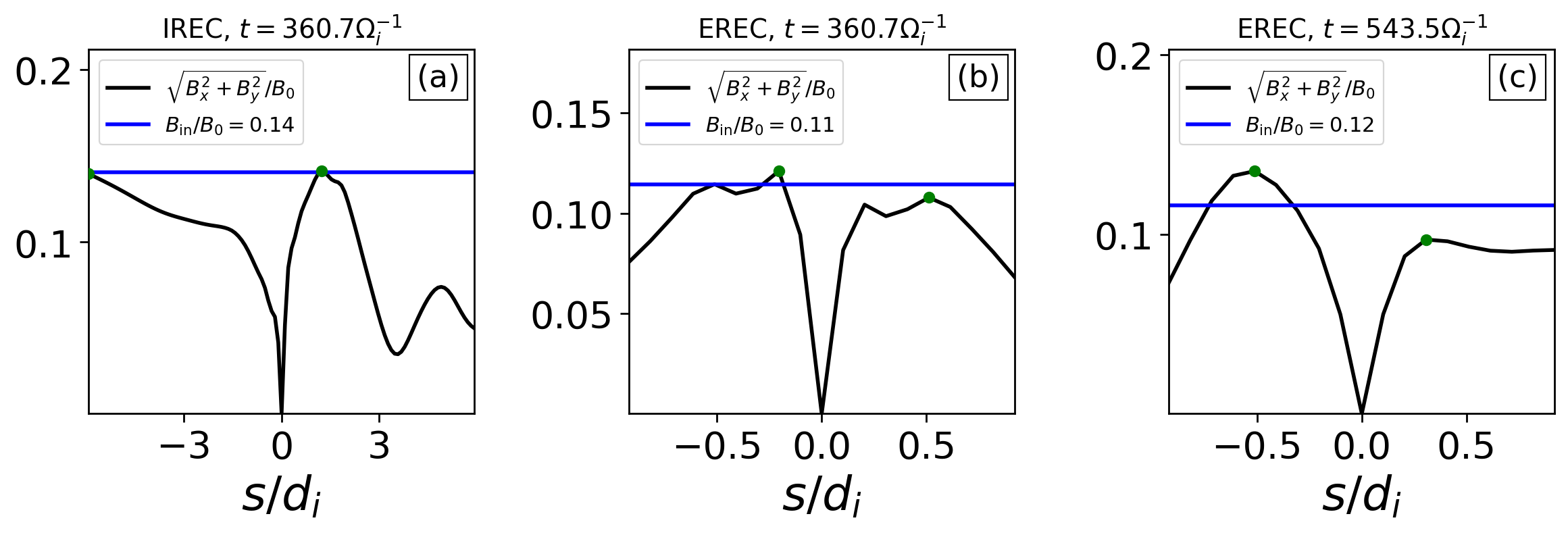}
\caption{Normalized in-plane magnetic field magnitude, $\sqrt{B_x^2+B_y^2}/B_0$, along a 1D cut through the inflow direction crossing the X-point for: (a) the IREC event at $t=360.7\Omega_i^{-1}$ (Fig. \ref{fig:FIG2_IREC}); (b) the secondary EREC event at $t=360.7\Omega_i^{-1}$ (Fig. \ref{fig:FIG3_EREC}); and (c) the EKHI-driven EREC event at $t=543.5\Omega_i^{-1}$ (Fig. \ref{fig:FIG6_EREC}). The green dots mark the local maxima for $s>0$ and $s<0$, while the blue line represents the average between these two maxima, defining $B_{\text{in}}/B_0$.}
\label{fig:appendix1}
\end{figure}

\begin{figure}[htbp]
\centering
\includegraphics[width=\linewidth]{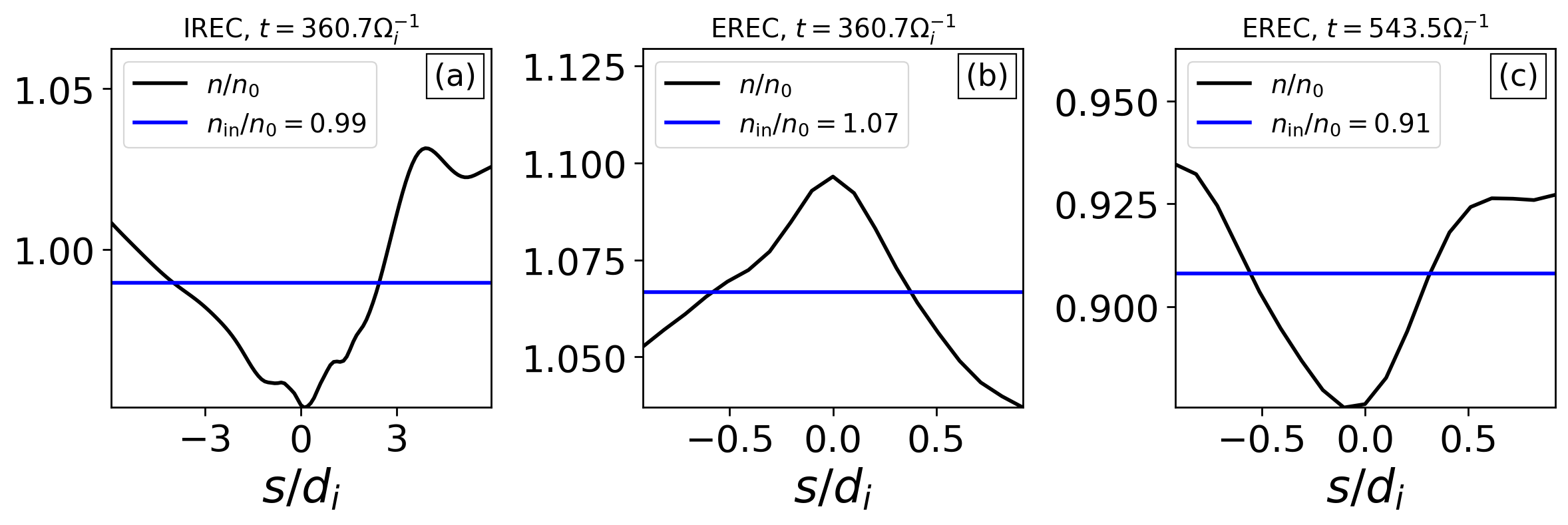}
\caption{Normalized density, $n/n_0$, along a 1D cut through the inflow direction crossing the X-point for: (a) the IREC event at $t=360.7\Omega_i^{-1}$ (Fig. \ref{fig:FIG2_IREC}); (b) the EREC event at $t=360.7\Omega_i^{-1}$ (Fig. \ref{fig:FIG3_EREC}); and (c) the EREC event at $t=543.5\Omega_i^{-1}$ (Fig. \ref{fig:FIG6_EREC}). The blue line represents the mean value along the cut, defining $n_{\text{in}}/n_0$.}
\label{fig:appendix2}
\end{figure}

In this appendix, we describe the calculation of the inflow Alfvén velocity, $c_{Ai,\text{in}} = B_{\text{in}}/\sqrt{4\pi m_i n_{\text{in}}}$, used to normalize the reconnection outflow velocities in Figures \ref{fig:FIG2_IREC}, \ref{fig:FIG3_EREC}, and \ref{fig:FIG6_EREC}. Here, $B_{\text{in}}$ and $n_{\text{in}}$ denote the magnitude of the inflow magnetic field and the inflow density, respectively. To quantify these parameters, we analyze a 1D cut along the inflow direction, defined as the eigenvector associated with the largest eigenvalue of the Hessian of the magnetic flux function at the X-point. The IREC and EREC cut lengths are set to $\sim 12d_i$ and $\sim 2d_i$, respectively, to appropriately match the width of the reconnection exhaust in both cases. $B_{\text{in}}$ is calculated as the average of the local magnetic field maxima found on both sides of the X-point (s=0), as illustrated in Figure \ref{fig:appendix1}. The inflow density $n_{\text{in}}$ is defined as the mean density along this 1D cut. As shown in Figure \ref{fig:appendix2}, $n_{\text{in}}$ remains close to the initial background density, $n_0$.

\section{Current sheet identification at early times}

\begin{figure}[htbp]
\centering
\includegraphics[width=\linewidth]{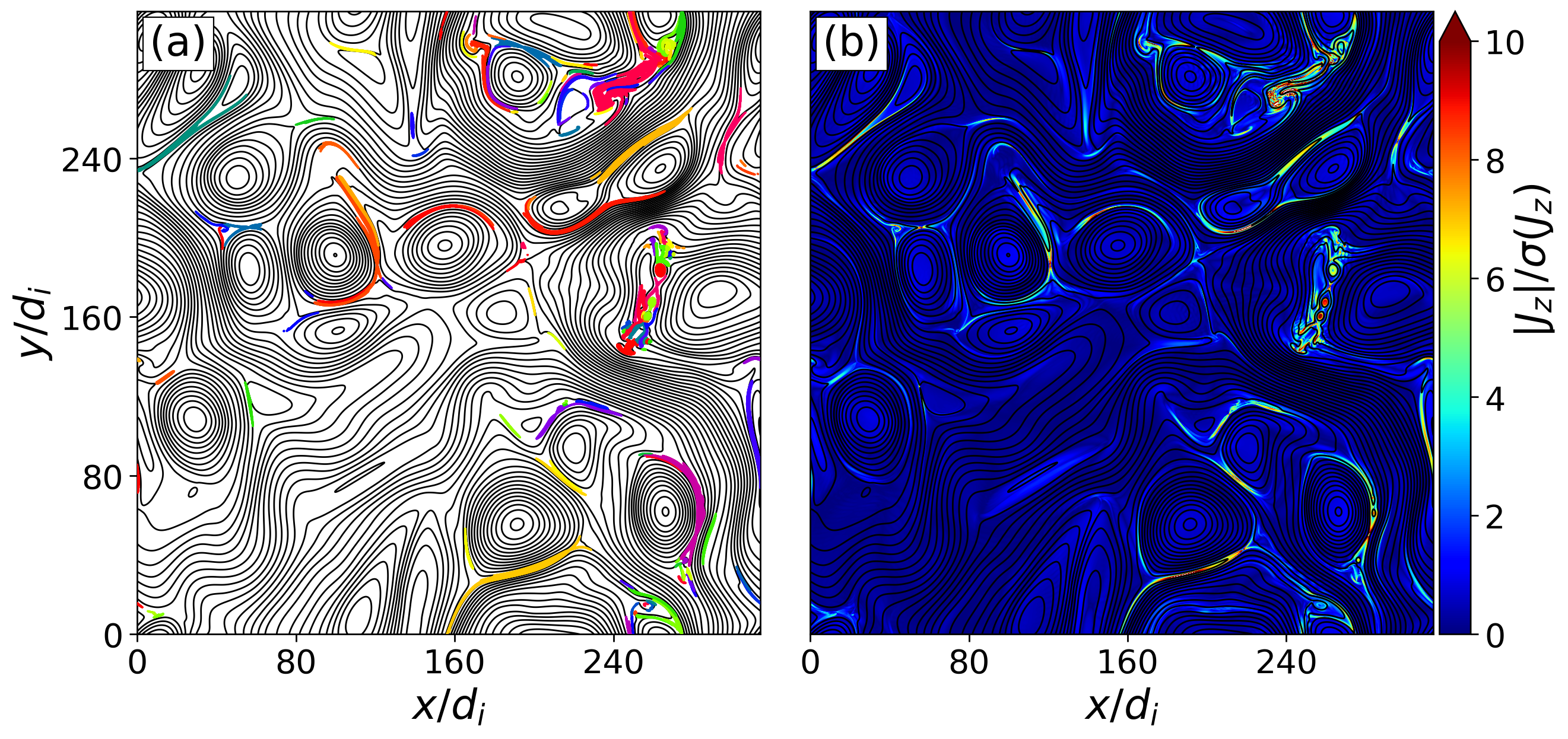}
\caption{Analysis of HDBSCAN clusters at time $t=320.7\Omega_i^{-1}$. Panel (a): In-plane magnetic field lines (black) with identified clusters highlighted in various colors. Panel (b): Shaded isocontours of the normalized out-of-plane current density $|J_z|/\sigma(J_z)$.}
\label{fig:appendix3}
\end{figure}

Figure \ref{fig:appendix3} displays the in-plane magnetic field lines at $t=320.7 \Omega_i^{-1}$, with identified clusters highlighted in various colors. At this time, the turbulent cascade has not yet fully developed and therefore does not reach very small scales. Consequently, HDBSCAN identifies a total of 143 clusters, demonstrating that at earlier times, CSs are fewer in number and larger in scale, as compared to the fully developed turbulent state where 516 CSs are detected.

\clearpage
\twocolumngrid

\bibliography{bibliography}

\end{document}